\documentclass[a4paper, oneside, twocolumn, notitlepage, 10pt]{extarticle_ecoc}
\usepackage{ecoc}
\usepackage{upgreek}
\usepackage{multirow}
\usepackage[position=top]{subfig}
\usepackage{tikz,pgfplots,subfig, filecontents}
\pgfplotsset{compat=1.18}
\usepackage{graphicx}
\usepackage{multicol}
\usepackage{colortbl}%
\usepackage{longtable}
\usepackage{caption}
\usepackage{fancyhdr}
\hypersetup{bookmarksdepth=-2}

\begin{document}
\selectlanguage{english}    

\title{Multi-Task Learning to Enhance Generalizability of\\ Neural Network Equalizers in
Coherent Optical Systems }%

\author{
     Sasipim Srivallapanondh\textsuperscript{(1)}, Pedro J. Freire\textsuperscript{(1)}, Ashraful Alam\textsuperscript{(1)}, Nelson Costa\textsuperscript{(2)}, Bernhard Spinnler\textsuperscript{(3)},\\   Antonio Napoli\textsuperscript{(3)}, Egor Sedov\textsuperscript{(1)}, Sergei K. Turitsyn\textsuperscript{(1)}, Jaroslaw E. Prilepsky\textsuperscript{(1)}
}
\maketitle                  
\thispagestyle{fancy}
\vspace{-3.5mm}
\begin{strip}
 \begin{author_descr}

   \textsuperscript{(1)}~Aston University, United Kingdom
   \textcolor{blue}{\uline{s.srivallapanondh@aston.ac.uk}}
   \textsuperscript{(2)}~Infinera, Carnaxide, Portugal 
   \textsuperscript{(3)}~Infinera, Munich, Germany\\
\vspace{-2.5mm}
 \end{author_descr}
 \vspace{-2.5mm}
\end{strip}
\vspace{-2.5mm}
\setstretch{1.1}

\vspace{-1mm}
\begin{strip}
\begin{ecoc_abstract}
    For the first time, multi-task learning is proposed to improve the flexibility of NN-based equalizers in coherent systems.  A ``single" NN-based equalizer improves Q-factor by up to 4~dB compared to CDC, without re-training, even with variations in launch power, symbol rate, or transmission distance.
\end{ecoc_abstract}
\vspace{-1mm}
\end{strip}
\vspace{-3mm}


\section{Introduction}
\vspace{-0.5mm}
The demand for high-speed data transmission keeps increasing due to upcoming technologies (6G\cite{holma2021extreme}, etc.). Coherent optical systems have emerged as a key solution to meet this demand. Nonetheless, the presence of linear and especially nonlinear distortions in fiber-optic  systems limits the achievable information rates \cite{essiambre2012capacity, cartledge2017digital, bayvel2016maximizing}. Various digital signal processing (DSP) techniques have been proposed for nonlinear effects mitigation in long-haul systems\cite{cartledge2017digital}. Neural networks (NNs) have recently emerged as an effective alternative for channel equalization: the NNs have demonstrated excellent capability to approximate the inverse of the optical channel transfer function, potentially outperforming conventional DSP approaches \cite{khan2019optical,hager2018nonlinear,freire2021performance}. However, generalizability remains one of the main challenges of NN-based equalizers and attracts more attention\cite{freire2021transfer,freire2022domain,zhang2022meta}. Due to different values of accumulated chromatic dispersion (CD) \cite{liu2023area}, or the presence of channel distortion, the equalizers in the receiver or transmitter require reconfiguration and must be adjustable to compensate for the variation of impairments as the channel characteristics change. 
 

In this work, multi-task learning (MTL) \cite{caruana1998multitask} is proposed to calibrate the NN-based equalizer used for different transmission conditions in coherent systems. MTL leverages shared representations to enhance the adaptability of NN-based equalizers across different system configurations and optical impairments.  This approach does not require re-training or additional data when the channel conditions change. Our results demonstrate the effectiveness of an MTL-based NN equalizer, which not only improves the equalization performance but also works efficiently in different transmission regimes and scenarios, leading to more generalizable and flexible solutions for NN-based nonlinear transmission effect mitigation.
 
\begin{figure*}[!ht]
\centering 
\vspace{-2mm}
  \includegraphics[width=\linewidth]{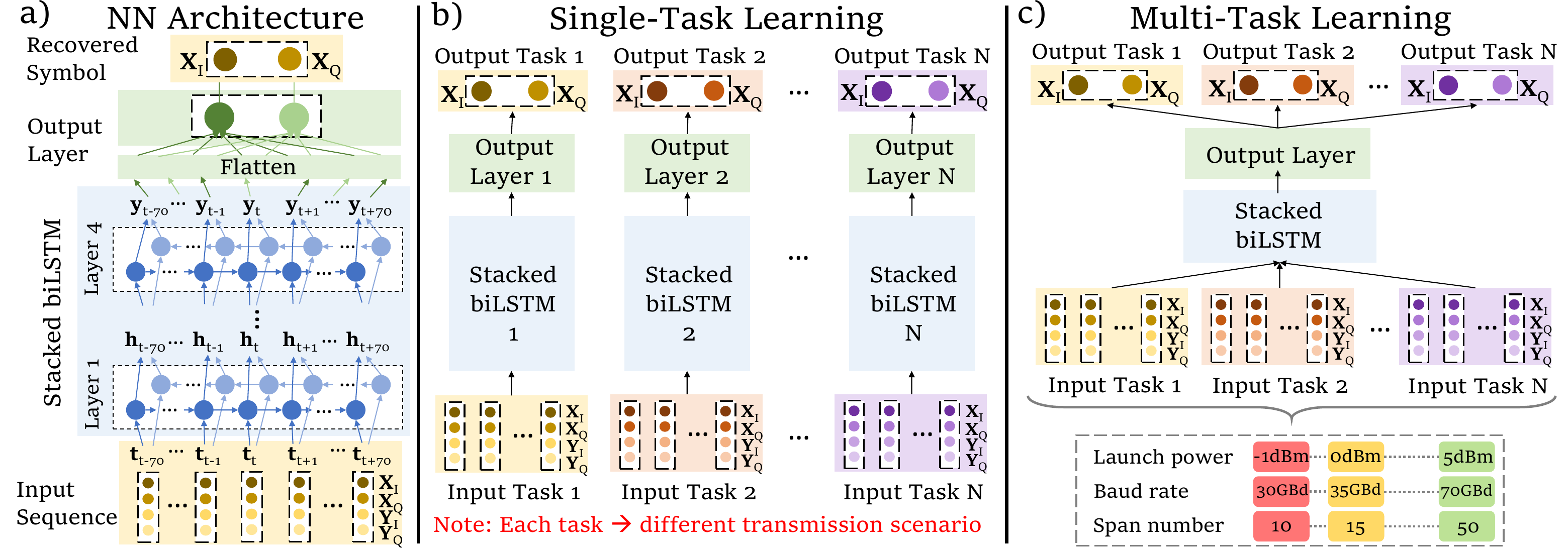}
  \caption{a) Equalizer architecture with 4-layer biLSTM and a dense layer; b) STL: multiple models are required for multiple transmission scenarios; c) MTL: only one  model is required for multiple transmission scenarios.}\label{fig:figure1}
  \vspace{-2mm}
\end{figure*}

\vspace{-1.6mm}

\section{Multi-Task Learning for NN-based Equalizers}
\vspace{-0.5mm}
Single Task Learning (STL) is a commonly used approach to train NNs. STL refers to the training in which the NN learns the representation of the function to provide the output of a ``specific'' task \cite{caruana1998multitask}. One advantage of STL is that it allows the NN to focus solely on a specific task, usually leading to very good performance in that task. However, the NN may behave poorly when applied to different tasks (e.g., when the transmission scenario of interest is not included in the initial training dataset). As shown in Fig.~\ref{fig:figure1}b, if STL is used for channel equalization in different transmission scenarios, multiple NN models are usually required to provide acceptable performance. In MTL, the NN is trained with multiple datasets from multiple related tasks. In this case, the common representations learned from different but related tasks are shared \cite{crawshaw2020multi, caruana1998multitask}. As depicted in Fig.~\ref{fig:figure1}c, MTL enables a single NN to equalize the signal in different ranges of launch power, symbol rate, and transmission distance by the joint training on the datasets from different transmission scenarios. MTL allows the NN to generalize better by using the domain-specific information contained in the different related tasks \cite{caruana1998multitask}.

Besides the generalization feature enabled by the MTL, it reduces hardware costs. In fact, the shared weights are fixed, which results in the simplification of the multipliers \cite{liu2023area}. However, MTL can also lead to some disadvantages compared to the STL. Firstly, there is a trade-off between the performance of individual tasks and the overall performance of the equalizer. Secondly, the degree of information sharing between tasks has to be carefully controlled. Too much sharing can cause a negative information transfer, resulting in performance degradation for each task \cite{crawshaw2020multi}.

In this work, we investigate the performance of NN-based equalizers using MTL where a single NN, without re-training, is potentially capable of recovering the transmitted symbol independently of the specific parameters of the transmission systems.
The considered transmission setup is altered by changing the symbol rate ($R_S$) and launch power ($P$) of data channels and the transmission distance (number of spans, $N_{Span}$). For the MTL, the NN is trained with different datasets resulting from the combination of different transmission setups (to share the weights and biases). 

\vspace{-2mm}
\section{Numerical Setup}
 \vspace{-0.5mm}
The dataset was obtained by numerical simulation assuming the transmission of a single 16-QAM dual-polarization channel along the standard single-mode fiber (SSMF). The signal propagation through the fiber was represented by a generalized Manakov equation using the GPU-accelerated split-step Fourier method\cite{esf0_2023_7880552}. The SSMF is characterized by the effective nonlinearity coefficient $\gamma$ = 1.2 (W$\cdot$ km)$^{-1}$, chromatic dispersion coefficient $D$ = 16.8 ps/(nm$\cdot$km), and attenuation parameter $\alpha$ = 0.21 dB/km. At the end of each fiber span, the optical fiber losses were compensated by an erbium-doped fiber amplifier with a noise figure of 4.5~dB. Downsampling and CD compensation (CDC) were performed on the receiver end. Afterwards, the received symbols were normalized and used as inputs of the NN.

\vspace{-2mm}
\section{Methodology}
\vspace{-0.5mm}
The NN architecture, depicted in Fig.~\ref{fig:figure1}a, contains a stack of four bidirectional-Long Short-Term Memory (biLSTM) layers with 100 hidden units in each layer coupled with a dense output layer of 2 neurons to deliver the real and imaginary values for the $X$-polarization. The biLSTM was selected because it outperformed other types of NNs when used for nonlinear compensation \cite{freire2021performance,deligiannidis2020compensation}. The model took four input features resulting from the in-phase and quadrature components of the complex signal ($X_I, X_Q, Y_I$, and $Y_Q$) where $X_I+jX_Q$ and $Y_I+jY_Q$ were the signals in the $X$ and $Y$ polarizations, respectively. A set of 141 input symbols was fed to the NN to recover one symbol at the output. A new set of synthetic data of size $2^{18}$ was randomly created with different system parameters and used in each training epoch to allow the model to learn different transmission scenarios. The entire training was carried out with a mini-batch size of 2000, and a learning rate of 0.001. The mean square error (MSE) loss estimator and the classical Adam algorithm ~\cite{gulli2017deep} were applied when training the weights and biases.

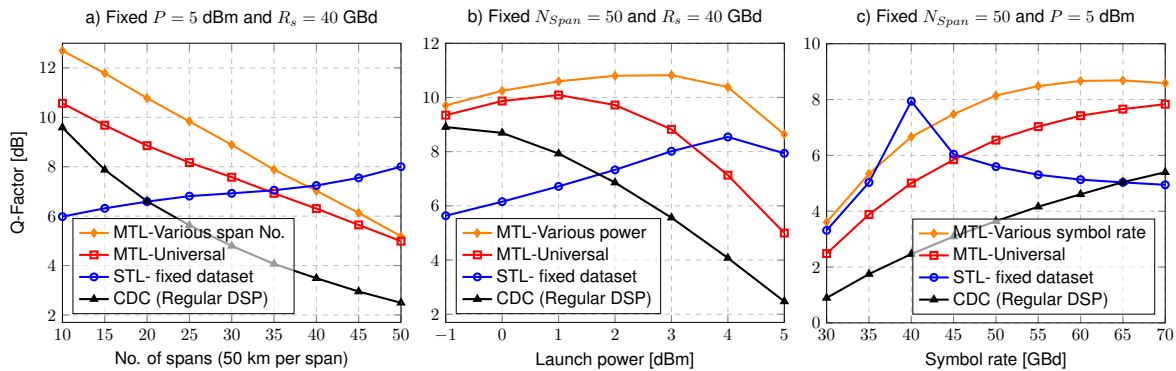
\begin{figure*}[!ht]
\vspace{-2mm}
\minipage{0.30\textwidth}
  \begin{tikzpicture}[scale=0.65]
    \begin{axis} [ 
        xlabel={No. of spans (50~km per span)},
        ylabel={Q-Factor [dB]},
        grid=both,  
        xmin=10, xmax=50,
    	xtick={10,15,20, ..., 50},
    	ymin=1.7, ymax=13,
        legend style={legend pos=south west, legend cell align=left,fill=white, fill opacity=0.6, draw opacity=1,text opacity=1},
        title = {a) Fixed $P = 5$~dBm and $R_s = 40$~GBd},
    	grid style={dashed}]
        ]
    \addplot[color=orange,mark=diamond, very thick]    coordinates {
    (10, 12.702)(15, 11.783)(20, 10.777)(25, 9.835)(30, 8.886)(35, 7.881)(40, 7.027)(45, 6.129)(50, 5.187)
    };
    \addlegendentry{MTL-Various span No.};
    
    \addplot[color=red,mark=square, very thick]    coordinates {
    (10, 10.565)(15, 9.681)(20, 8.853)(25, 8.169)(30, 7.577)(35, 6.926)(40, 6.306)(45, 5.645)(50, 4.986)
    };
    \addlegendentry{MTL-Universal};

    \addplot[color=blue,mark=o, very thick]    coordinates {
    (10, 5.982)(15, 6.316)(20, 6.588)(25, 6.811)(30, 6.923)(35, 7.049)(40, 7.24)(45, 7.549)(50, 8.002)
    };
    \addlegendentry{STL- fixed dataset};
    
    \addplot[color=black,mark=triangle, very thick]    coordinates {
    (10, 9.585)(15, 7.873)(20, 6.601)(25, 5.628)(30, 4.794)(35, 4.066)(40, 3.488)(45, 2.948)(50, 2.492)
    };
    \addlegendentry{CDC (Regular DSP)};
    \end{axis}
    \end{tikzpicture}
    \label{fig:plot1}
\endminipage\hfill
\minipage{0.30\textwidth}
  \begin{tikzpicture}[scale=0.65]
    \begin{axis} [ 
        xlabel={Launch power [dBm]},
        grid=both,  
        xmin=-1, xmax=5,
    	xtick={-1, ..., 5},
    	ymin=1.7, ymax=12,
        legend style={legend pos=south west, legend cell align=left,fill=white, fill opacity=0.6, draw opacity=1,text opacity=1},
        title = {b) Fixed $N_{Span} = 50$ and $R_s = 40$~GBd},
    	grid style={dashed}]
        ]
    \addplot[color=orange,mark=diamond, very thick]    coordinates {
    (-1, 9.707)(0, 10.246)(1, 10.595)(2, 10.801)(3, 10.822)(4, 10.383)(5, 8.638)
    };
    \addlegendentry{MTL-Various power};
    
    \addplot[color=red,mark=square, very thick]    coordinates {
    (-1, 9.346)(0, 9.868)(1, 10.09)(2, 9.72)(3, 8.823)(4, 7.132)(5, 4.997)
    };
    \addlegendentry{MTL-Universal};

    \addplot[color=blue,mark=o, very thick]    coordinates {
    (-1, 5.635)(0, 6.153)(1, 6.718)(2, 7.329)(3, 8.015)(4, 8.542)(5, 7.938)
    };
    \addlegendentry{STL- fixed dataset};
    
    \addplot[color=black,mark=triangle, very thick]    coordinates {
    (-1, 8.908)(0, 8.692)(1, 7.929)(2, 6.859)(3, 5.563)(4, 4.075)(5, 2.472)
    };
    \addlegendentry{CDC (Regular DSP)};
    \end{axis}
    \end{tikzpicture}
    \label{fig:plot2}
\endminipage\hfill
\hspace*{-0.5cm}
\minipage{0.31\textwidth}
   \begin{tikzpicture}[scale=0.65]
    \begin{axis} [ 
        xlabel={Symbol rate [GBd]},
        grid=both,  
        xmin=30, xmax=70,
    	xtick={30, 35, 40, ..., 70},
    	ymin=0, ymax=10,
        legend style={legend pos=south east, legend cell align=left,fill=white, fill opacity=0.6, draw opacity=1,text opacity=1},
        title = {c) Fixed $N_{Span} = 50$ and $P = 5$~dBm},
    	grid style={dashed}]
        ]
    \addplot[color=orange,mark=diamond, very thick]    coordinates {
    (30, 3.6089)(35, 5.3399)(40, 6.6627)(45, 7.4764)(50, 8.1446)(55, 8.481)(60, 8.6685)(65,  8.6863)(70,8.5888)
    };
    \addlegendentry{MTL-Various symbol rate};

    \addplot[color=red,mark=square, very thick]    coordinates {
    (30, 2.483)(35, 3.883)(40, 5.009)(45, 5.846)(50, 6.548)(55, 7.031)(60, 7.422)(65, 7.658)(70, 7.834)
    };
    \addlegendentry{MTL-Universal};

    \addplot[color=blue,mark=o, very thick]    coordinates {
    (30, 3.315)(35, 5.031)(40, 7.942)(45, 6.041)(50, 5.598)(55, 5.306)(60, 5.132)(65, 5.032)(70, 4.946)
    };
    \addlegendentry{STL- fixed dataset};
    
    \addplot[color=black,mark=triangle, very thick]    coordinates {
    (30, 0.894)(35, 1.748)(40, 2.465)(45, 3.093)(50, 3.644)(55, 4.162)(60, 4.606)(65, 5.04)(70, 5.397)
    };
    \addlegendentry{CDC (Regular DSP)};
    \end{axis}
    \end{tikzpicture}
    \label{fig:plot3}
\endminipage\hfill
%
  \vspace{-4mm}
\caption{Q-factor resulting from using MTL (orange and red) and STL model (blue) in the following test cases; a) when the transmission distance changes but the launch power and symbol rate are set to 5~dBm and 40~GBd, respectively; b) when the launch power changes but the number of span and symbol rate are set to 50 and 40~GBd, respectively; c) when the symbol rate changes but the number of spans and launch power are set to 50 and  5~dBm, respectively.}
\vspace{-1mm}
\label{fig:all_plots}
\end{figure*}

The transmission scenarios include $R_S$ ranging from 30 to 70~GBd, number of spans ranging between 10 and 50 (with fixed 50~km span length), and launch power ranging between -1 and 5~dBm. The NNs were trained with MTL or STL as follows: 

\begin{enumerate}
    \item MTL trained for 1000 epochs with datasets including different $N_{Span}$, but fixed $R_S$ = 40~GBd and $P$~=~5~dBm.
    \item MTL trained for 1000 epochs with datasets including different $P$, but fixed $N_{Span}$~=~50 and $R_S$~=~40~GBd\footnote{This model has one extra input feature, which is the launch power. The model learns the data during the training using a normalized launch power. Therefore, it could not learn to generalize well without knowing the actual launch power.}.
    \item MTL trained for 1000 epochs with datasets including different $R_S$ but fixed $N_{Span}$ = 50 and $P$ = 5~dBm.
    \item MTL trained for 1200 epochs with datasets including different combinations of $N_{Span}$, $R_S$, and $P$. This NN is referred to as the ``Universal model''\footnote{Here, the values of $R_S$ and $N_{Span}$ are randomly selected from the list of possible baud rate values with 5~GBd increment and the list of span number with the increments of 5 spans, respectively, to decrease the possible number of combinations for the NN's learning.}.
    \item STL (without MTL) trained for 1000 epochs with fixed parameters: $R_S$ =~40~GBd, $N_{Span}$~=~50 and $P$~=~5~dBm.
\end{enumerate}

\vspace{-3mm}
\section{Results and Discussion}
\vspace{-1mm}
We considered MTL for multiple symbol rates, transmission distances, and launch powers. To evaluate equalization performance and generalizability, the MTL models were compared to  CDC and the STL model trained with a fixed dataset. 

\underline{Variation of transmission distance:} Fig.~\ref{fig:all_plots}a shows the optical performance for different reaches considering a fixed launch power of 5~dBm and a signal baud rate of 40~GBd. The STL model performed the best when $N_{Span}$ was 50 (because it was trained for this specific transmission scenario), significantly outperforming the remaining approaches. However, its performance was significantly impacted in the shorter reaches as it could not generalize. On the other hand, the MTL trained with different $N_{span}$ showed much better performance than STL for the shorter reaches, achieving a better Q-factor (about 3~dB Q-factor improvement) than CDC only for all considered scenarios. The  universal MTL model also showed better performance than the CDC alone, leading to a maximum Q-factor improvement of about 2.5~dB  at 50$\times$50~km.

\underline{Variation of launch powers: } Fig.~\ref{fig:all_plots}b depicts the Q-factor as a function of the launch power for a fixed $R_S$ of 40~GBd and transmission distance of 50$\times$50~km.
Again, the STL model showed the best gain for launch powers close to the one it was trained with (5~dBm), but revealed quite poor results for the remaining launch powers. In contrast, the universal MTL model enabled a Q-factor improvement exceeding 2~dB for the most relevant launch powers. The MTL, trained with various $P$ but fixed $N_{SPAN}$ and $R_S$, revealed the best performance, enabling a Q-factor improvement exceeding 4~dB for the most relevant launch powers. Interestingly, we can see that, at 5~dBm, the MTL outperformed STL. The reason for this may be that the STL is overfitting and cannot adapt to the unseen test data as effectively as the MTL model, which is more generalized. Ref. \cite{srivallapanondh2022knowledge} supported the claim that a more generalized model can perform better.

\underline{Variation of symbol rates:} Fig.~\ref{fig:all_plots}c illustrates the Q-factor as a function of the data signal baud rate for a fixed transmission distance and launch power of 50$\times$50~km and 5~dBm, respectively. STL led to very good results for the 40~GBd transmission scenario (training scenario) but showed very poor generalization capability. The MTL, trained with multiple $R_S$ but fixed $N_{Span}$ and $P$, enabled a Q-factor improvement of up to 4.5~dB with respect to the CDC only, whereas the universal MTL model showed up to 2.5~dB improvement. The MTL provided a good gain in most cases.

The aforementioned results 
show that, although STL may lead to outstanding performance in specific transmission conditions, it is not suitable for real-world system application because it lacks the adaptability to dynamic optical network parameters. MTL overcomes this limitation, allowing the equalizer to be more flexible, but at the cost of small performance degradation compared to models trained only for a specific task. 


\vspace{-2.5mm}
\section{Conclusions}
Multi-task learning is proposed to allow a ``single'' NN-based equalizer, without re-training, to recover received symbols when the transmission scenarios change. The results showed that the MTL can provide up to 4~dB improvement in Q-factor with respect to CDC alone even if the transmission distance, launch power, and symbol rate vary, thus highlighting the adaptability of the MTL NN-based equalizer to the real-world dynamic optical network.

\vspace{0.5mm}
\footnotesize
\linespread{0.0}
\textbf{Acknowledgements:} This work is supported by the EU H2020 Marie Skodowska-Curie Action project MENTOR (No. 956713), SMARTNET EMJMD program under Grant 586686-EPP-1-2017-1-U.K.-EPPKA1-JMDMOB, EPSRC project TRANSNET (EP/R035342/1), and the Horizon Europe project ALLEGRO, GA n. 101092766.
\normalsize
\linespread{1.0}
\printbibliography
\end{document}